# EpiBeds: Data informed modelling of the COVID-19 hospital burden in England

Christopher E. Overton[*,1,2,3], Lorenzo Pellis[*,1,3,14], Helena B. Stage[1,4,5], Francesca Scarabel[1,3], Joshua Burton[7], Christophe Fraser[9,11,12], Ian Hall[1,2,3,14,15], Thomas A. House[1,2,3,7,14], Chris Jewell[8], Anel Nurtay[9], Filippo Pagani[1,6], Katrina A. Lythgoe[9,10]

*These authors contributed equally.
[1] Department of Mathematics, University of Manchester, UK.
[2] Clinical Data Science Unit, Manchester University NHS Foundation Trust, UK.
[3] Joint UNIversities Pandemic and Epidemiological Research, https://maths.org/juniper/.
[4] The Humboldt University of Berlin, Germany.
[5] The University of Potsdam, Germany.
[6] MRC Biostatistics Unit, University of Cambridge, UK.
[7] Division of Informatics, Imaging and Data Sciences, Faculty of Biology Medicine and Health, University of Manchester, UK.
[8] CHICAS, Lancaster Medical School, Lancaster University, UK.
[9] Big Data Institute, Nuffield Department of Medicine, University of Oxford, UK.
[10] Department of Zoology, University of Oxford, UK.
[11] Wellcome Centre for Human Genetics, Nuffield Department of Medicine, NIHR Biomedical Research Centre, University of Oxford, UK.
[12] Wellcome Sanger Institute, Cambridge, UK.
[13] IBM Research, Hartree Centre, Daresbury, UK.
[14] Alan Turing Institute, UK.
[15] Emergency Preparedness, Health Protection Division, Public Health England, UK.

**Abstract**
The first year of the COVID-19 pandemic put considerable strain on the national healthcare system in England. In order to predict the effect of the local epidemic on hospital capacity in England, we used a variety of data streams to inform the construction and parameterisation of a hospital progression model, which was coupled to a model of the generalised epidemic. We named this model EpiBeds. Data from a partially complete patient-pathway line-list was used to provide initial estimates of the mean duration that individuals spend in the different hospital compartments. We then fitted EpiBeds using complete data on hospital occupancy and hospital deaths, enabling estimation of the proportion of individuals that follow different clinical pathways, and the reproduction number of the generalised epidemic. The construction of EpiBeds makes it straightforward to adapt to different patient pathways and settings beyond England. As part of the UK response to the pandemic, EpiBeds has provided weekly forecasts to the NHS for hospital bed occupancy and admissions in England, Wales, Scotland, and Northern Ireland.

## 1. Introduction

SARS-CoV-2 was first discovered in China in late 2019, and subsequently seeded epidemics throughout the world. The UK-wide lockdown implemented on 23rd March 2020 suppressed the initial surge in infections, but the subsequent easing of physical distancing interventions led to further surges both regionally and nationally, requiring the introduction of subsequent restrictions and lockdowns. As part of the UK response, we developed a short-term forecast model for predicting hospital and critical care bed use, which was combined with modelling work from other research groups to inform the National Health Service (NHS) of future COVID-19 bed demand.

Due to the potentially fast-growing nature of COVID-19 outbreaks [1], and the short duration of most interventions, long-term forecasting is limited, since conditions are likely to have changed between the production of the forecast and reaching the forecast horizon. Because of the delay of a few weeks between the implementation of interventions and their effects on hospital admissions [1,2], short-term forecasts of a few weeks are not significantly affected (except around known major control policy changes, e.g. lockdowns), and are valuable planning tools for hospitals and health services.

We developed a minimally complex compartmental model tailored to data available on hospital flow and our knowledge of the natural history of disease progression. Specifically, hospital compartments were defined by both the current status of the patient (e.g. in critical care) and their outcome (e.g. will recover), since outcome is a major determinant of durations of stay in different compartments [3]. Additionally, we modelled a simplified community epidemic to estimate the flow of individuals into the hospital system using other studies to parameterise the natural history of disease progression outside of a hospital setting. We fitted EpiBeds to four data streams: daily hospital admissions, daily hospital prevalence, daily intensive care unit (ICU) prevalence, and daily deaths in hospital.

We reduced the number of unknown parameters using high-resolution individual-level data for a subset of hospitalised patients in England, to estimate the length of stay in each hospital compartment of the EpiBeds model. The estimates are valuable results by themselves, as although a low incidence rate of hospitalisation may suggest limited pressure on the health care system, if the length of time in hospital is large, or the ICU admission rate is high, the prevalence in hospital can present a substantial burden. Other outputs of EpiBeds, beyond the short-term forecasts, include the probabilities of moving to different hospital compartments (which gives a measure of the severity of hospitalisation with COVID-19).

Since hospitalisation data reflect background incidence (though the relationship can change over time with changing age profile of cases or vaccination, for example), EpiBeds also allows us to approximate the growth and transmission rates in the background epidemic, and hence the reproduction number. We consider two types of reproduction number: the control reproduction number $R_c(t)$ and the effective reproduction number $R_e(t)$ (often denoted $Rt$). In short, $R_c(t)$ describes the reproduction number at time t in the absence of population immunity, but in the presence of other control interventions (such as physical distancing), whereas $R_e(t)$ in addition takes account of population immunity [4].

The original motivation for the development of EpiBeds was to provide real-time estimates of the instantaneous growth rate and effective reproduction number, published weekly by the UK Government, as well as short-term projections of hospital and ICU bed demand to support the resource management of the National Health Service of England, nationally and separately for each region, and the other Devolved Administrations in the UK. Weekly estimates and projections from a suite of different models, all developed independently and fit with different methods to a variety of data streams, were submitted to the Scientific Pandemic Influenza Group on Modelling (SPI-M) and combined by CrystalCast [5] to provide the most robust real-time indicators of current epidemic trends. When policy was known to have changed recently or to be about to change, often multiple scenarios were submitted in addition to the projections (which assumed no change in transmission from the day of the projection), with a range of fixed values for the reproduction number from the date of the policy change.

The relative simplicity of EpiBeds makes it more transparent than more complex models [4,6,7], and unlike other models enables us to estimate the probability of moving between different hospital compartments. Although developed for England, it can be applied to other countries/regions by estimating appropriate length of stay distributions, if such data are available. However, the flexibility of its construction and parameterisation means it can easily be adapted to provide accurate short-term forecasts for different countries and healthcare systems, with the model structure tailored to the observed data.

## 2. Results

First, to inform the EpiBeds model structure, we analysed the detailed COVID-19 Hospitalisation in England Surveillance System (CHESS) and Severe Acute Respiratory Infection (SARI) datasets (see Section M.1) to identify the most relevant hospital pathways and to estimate the distributions of the time individuals spent along each step of these hospital pathways. Using the results, together with disease parameters from the literature, we constructed a minimally complex ordinary differential equation (ODE) compartmental model of SARS-CoV-2 transmission with a detailed model component for the flow of COVID-19 patients through hospitals. By fitting EpiBeds to four data streams (hospital admissions, hospital prevalence, ICU prevalence and hospital deaths; see Section M.1), we derived severity estimates, i.e. estimates of the probability of each possible patient outcome. We report these separately for the first and second epidemic waves, along with the history of the reproduction number at different times during the pandemic. We then show how EpiBeds was used for short-term projections of the hospital and ICU bed demand, together with real-time estimates of the effective reproduction number $R_e(t)$. Here, we present such estimates twice a month, with an assessment of the quality of the projections.

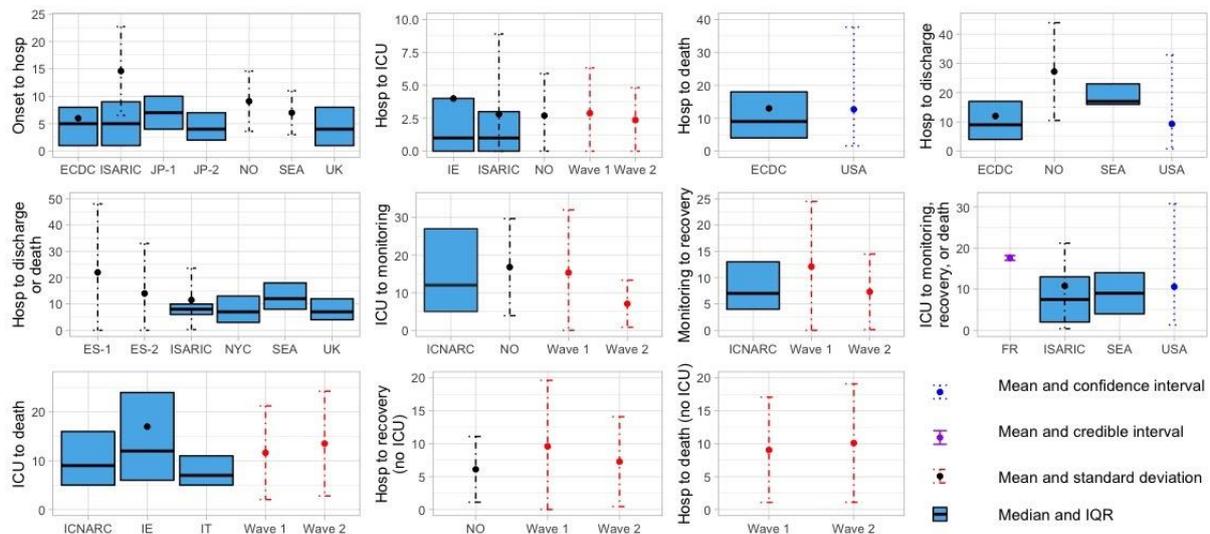

**Figure 1:** *Length of stay and delay distributions relating to hospitalisations in Europe and North America. Estimates of various lengths of delays prior to hospital admissions, and the lengths of stay through hospital patient pathways. Our estimates (red, labelled Wave 1 and Wave 2 for each waves) derive from the CHESS data, with other sources including aggregated values from the European Centre for Disease Prevention and Control (ECDC, [8]), ISARIC reports (ISARIC, [9]), the Norwegian Institute of Public Health (NO, [10]), a study among Seattle patients (SEA, [11]), ISARIC patients in the UK (UK, [12]), a study of severely ill patients in New York (NYC, [13]), Irish reported data to the ECDC (IE, [8]), ICNARC reports (ICNARC, [14]), a study on ICU patients in Lombardy (IT, [15]), a cohort study in California and Washington (USA, [16]), Japanese hospital patients (JP, [17]), Spanish hospital patients in Reus (ES, [18]), and French hospitalisation data (FR, [19]). If not specified, all estimates are from the first wave. Otherwise, estimates are labelled for each wave but may not agree on the respective date ranges. Mean values are marked by dots, and median values and interquartile ranges by box plots. All error bars denote standard deviations (dot-dashed), except for FR where these bounds denote the 95% credible interval (solid), and USA where the bounds denote the 95% confidence interval (dotted), and therefore cannot be directly compared. There is a marked skew in several distributions as seen by the difference between mean and median values. We do not include estimates from China due to documented evidence that these distributions differ [20,21].*

## 2.1 Estimates of hospital length of stay distributions

From the CHESS/SARI dataset, we first identified the possible status of hospital patients, as well as the observed transitions between states. Based on the most relevant transitions, we classed patients using five states: Hospitalised (not been to ICU), in Critical care (ICU), Monitored (discharged from

ICU but still in hospital), Recovered, and Deceased. After hospital admission, patients are either discharged, admitted to ICU, or die (without entering ICU), and from ICU individuals may go on to be discharged from ICU (but remain in hospital in the monitored state) or die. There are other potential pathways: for example, some patients who are discharged from ICU may re-enter ICU. However, these pathways have a relatively minor contribution compared to those outlined above, with insufficient data to accurately estimate the relevant parameters. Therefore, we omit these minor pathways from the model framework. We then estimated the distributions of the time individuals take for each transition (hereafter referred to as "length of stay" or "delay distribution", with the former preferred for in-hospital events and the latter preferred for out-of-hospital events), in particular: hospital admission to ICU admission, ICU admission to ICU discharge, ICU admission to death, ICU discharge to hospital discharge, hospital admission to death and hospital admission to hospital discharge. For hospital admission to death and hospital admission to hospital discharge, we only consider patients who are not admitted to ICU, to prevent overlap with the ICU-related pathways.

With the foresight of formulating the model as a set of ODEs, with constant rates between compartments, we assumed that the lengths of stay were Gamma-distributed. To keep EpiBeds as simple as possible, we did not model any covariates besides calendar time (see Section M.2). Calendar time is vital to consider, since several conditions are likely to change over time, for instance the demographic of hospital admissions or treatment policies (both of which can change patient outcomes and lengths of stay). Motivated by this observation, we generated monthly cumulative estimates, starting from two different dates to distinguish the first (from 01/03/2020 to the last day of each month, until 31/08/2020) and second (from 01/08/2020 to the last day of each month, until 31/12/2020) wave. The resulting estimated distributions obtained for both waves are collected in Table 1, with the monthly cumulative estimates given in Table A1. Our estimates are consistent with previous results for length-of-stay distributions, particularly findings for the UK (see Figure 1).

**Table 1.** *Gamma distributed length of stay for different events in hospital, estimated using the CHESS/SARI data. Brackets indicate 95% confidence intervals (generated through parametric bootstrapping).*

| Length of stay | Date range | Mean | Standard deviation | N |
| --- | --- | --- | --- | --- |
| **Hosp to ICU** | 01/03/2020 - 31/08/2020 | 2.79 (2.71, 2.87) | 3.30 (3.20, 3.41) | 6254 |
| | 01/08/2020 - 31/12/2020 | 2.70 (2.61, 2.79) | 2.96 (2.83, 3.07) | 3830 |
| **ICU to death** | 01/03/2020 - 31/08/2020 | 11.84 (11.43, 12.25) | 9.74 (9.35, 10.20) | 2268 |
| | 01/08/2020 - 31/12/2020 | 15.33 (14.50, 16.08) | 12.38 (11.52, 13.18) | 837 |
| **ICU to monitoring** | 01/03/2020 - 31/08/2020 | 15.93 (15.39, 16.52) | 16.97 (16.30, 17.64) | 3642 |
| | 01/08/2020 - 31/12/2020 | 8.57 (8.18, 8.98) | 7.51 (7.05, 7.96) | 1348 |
| **Monitoring to recovery** | 01/03/2020 - 31/08/2020 | 11.85 (11.39, 12.29) | 11.93 (11.37, 12.44) | 2602 |

|                          | 01/08/2020 - 31/12/2020 | 6.45 (6.04, 6.90)   | 6.58 (6.09, 7.09)   | 945  |
|--------------------------|-------------------------|---------------------|---------------------|------|
| **Hosp to recovery (no ICU)** | 01/03/2020 - 31/08/2020 | 9.37 (9.14, 9.60)   | 9.68 (9.41, 9.96)   | 6312 |
|                          | 01/08/2020 - 31/12/2020 | 10.02 (9.66, 10.42) | 9.89 (9.43, 10.35)  | 2462 |
| **Hosp to death (no ICU)** | 01/03/2020 - 31/08/2020 | 8.93 (8.58, 9.27)   | 7.81 (7.44, 8.16)   | 2144 |
|                          | 01/08/2020 - 31/12/2020 | 12.16 (11.43, 12.92)| 10.40 (9.59, 11.23) | 674  |

Comparing the first wave to the second wave, we observe substantial changes in the lengths of stay on ICU. The length of stay from entering ICU to dying slightly increased, whilst the length of stay from entering ICU to leaving ICU decreased by a factor of two. Similarly, the length of stay from leaving ICU to discharge decreased by a factor of two. There are various potential drivers for this. First, treatment changes could have reduced the length of time patients require critical care treatment, and prolonged the time until death. Second, younger patients, who were more common in the second wave, take less time to recover and longer to die. The lengths of stay without ICU does not show the same drop in the time to recovery as seen in ICU, but has a similar increase in the time to death, possibly because of improved quality of treatment.

*2.2 A compartmental model informed by hospital flow data*

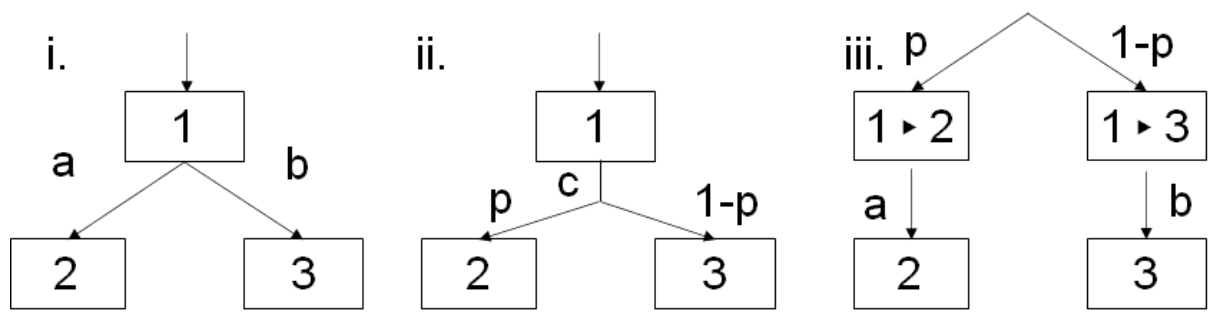

**Figure 2:** *Illustration of the different mechanisms for modelling transitions between compartments: (i.) two constant rates, a and b, to progress from state 1 to states 2 and 3, respectively; (ii.) a single rate c for leaving state 1, with progression to state 2 with probability p and to state 3 with the opposite probability; (iii.) the approach used in this work, i.e. state 1 split in two different states, based on the probability of eventual outcome state (p or its opposite for states 2 or 3 respectively), coupled with two separate rates, a and b, for the transition to the next state. Note that while (ii.) is equivalent to (i.) for c = a+b and p = a/(a+b), (iii.) differs from the other two, except in the special case of a = b = c.*

Informed by the estimated length of stay distributions (Table 1), we constructed a compartmental model describing the progression of individuals through the hospital pathways. To account for considerable differences in the duration of different hospital transitions even from the same state, we divided individuals into compartments both in terms of their current status (e.g., Hospitalised or

Critical care) and in terms of their future outcome (e.g., will recover, will die). This approach requires more parameters than the more common approach based on competing hazards (see Figure 2) but is more flexible (resulting in more general phase-type sojourn times in each state), can be directly parameterised with available data, and has the advantage of separating the temporal information related to conceptually different processes (see below). From the pathways identified in Section 2.1, we therefore considered three hospital admission compartments: $H_R$ (hospitalised - will recover), $H_C$ (hospitalised - will enter critical care), $H_D$ (hospitalised - will die without entering critical care), and similarly two critical care compartments: $C_M$ (critical care - will enter the monitoring compartment $M_R$ before recovery), and $C_D$ (critical care - will die). By fitting Gamma distributions to the lengths of stay, we observe that the standard deviation is close to the mean (Table 1) for all the transitions out of the states above, implying these Gamma distributions are approximately exponential (shape parameter 1). This means that the flows between hospital compartments are suitably described by constant transition rates (equal to the inverse of the mean of the exponentially distributed sojourn time in the compartment, see Section M.3 [22]). The resultant hospital flow is shown by the red and orange compartments in Figure 3.

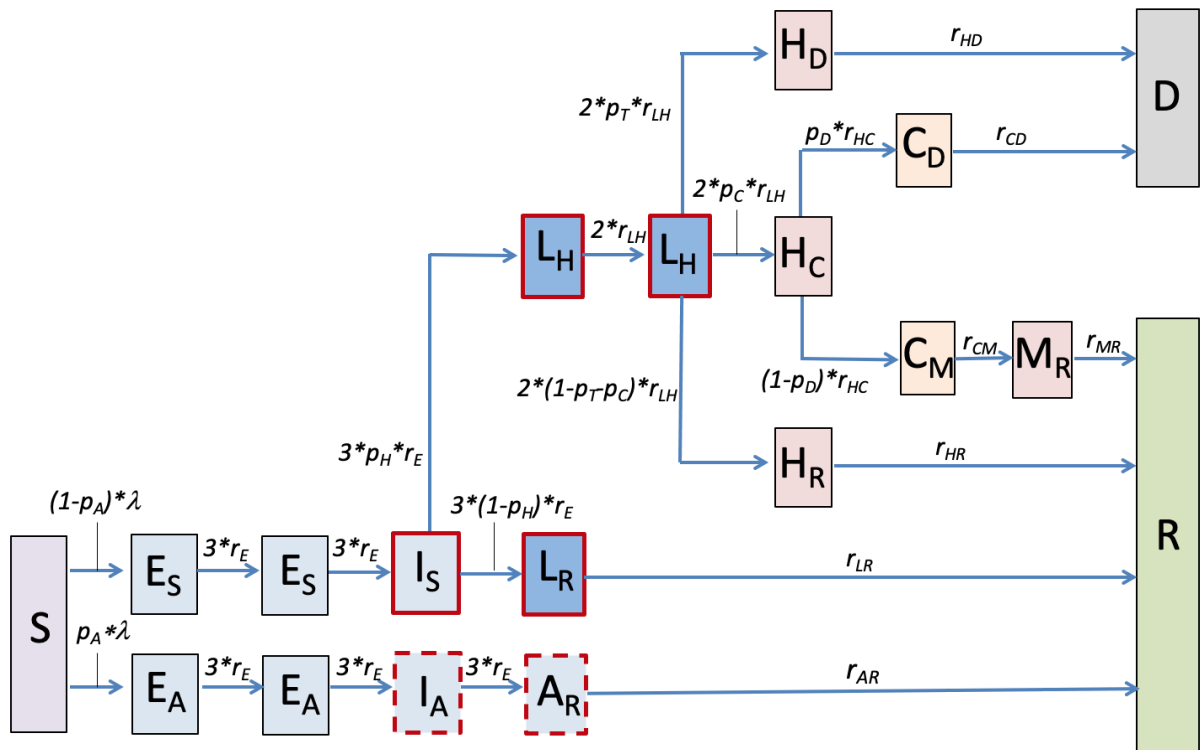

**Figure 3: Schematic representation of the EpiBeds model.** *The construction of this compartmental model was informed by available data. EpiBeds is implemented as a set of ODEs, with one state variable per compartment representing the absolute number of individuals in it. Arrows describe flow between compartments, which occurs at constant rate. Blue compartments indicate infected individuals who are not hospitalised, with a dark and light blue distinction, respectively, for individuals with and without symptoms, while red compartments indicate hospitalised individuals and orange compartments individuals in critical care. The compartments with a red border contain infectious individuals, with a dashed border denoting an infectivity reduced to 25% of that of the other infectious compartments; once hospitalised, it is assumed individuals no longer contribute to the community epidemic. For states in which the waiting times are not exponentially distributed (e.g. Exposed) we use multiple compartments enabling us to approximate Gamma-distributed waiting times. All variables, rates, and probabilities are described in Tables 2a and 2b. The force of infection $\lambda$ depends on the numbers in the infectious compartments, and its expression is reported in Section M.3.*

Since the infectious burden in the population determines the rate at which cases will be admitted to hospital, we also used compartments to describe the process of infection in the general population, based on an SEIR (Susceptible Exposed Infectious Recovered) model structure. Symptomatic individuals therefore go through three states of infection, Exposed (but not yet infectious), Infectious (but not yet symptomatic), and Late infection (infectious and symptomatic), with a proportion of symptomatic individuals requiring hospitalisation ($L_H$) and the other proportion recovering naturally ($L_R$). The latter distinction is motivated by the fact that the processes of biological recovery and hospital seeking behaviour are conceptually different, hence involving different progression rates: for an infected individual, the time to recovery reflects the natural course of a non-severe infection, while the time to hospital admission is driven by hospital seeking behaviour, current policy, and health-care logistic availability. A proportion of individuals are assumed to remain asymptomatic throughout infection; these individuals follow an infection pathway that is distinct from, but mimics, that of symptomatic individuals.

The structure for the generalised epidemic was constructed to reflect delay distributions from the literature, using constant rates to represent exponentially distributed permanence times, and sequences of compartments to represent Gamma-distributed permanence times (more details in Section M.3). Hence, to describe a Gamma-distributed incubation period (i.e., the time from infection to symptom onset) with mean 4.85 days and shape parameter three [1], we used three subsequent compartments ($E_1, E_2, I$) with identical constant rates between them, with mean waiting time 1.6 days in each compartment [22]. This assumes pre-symptomatic transmission of 1.6 days, which is roughly consistent with literature estimates that show most pre-symptomatic transmission occurs in the two days prior to symptom onset [23]. The delay between symptom onset and hospitalisation is Gamma distributed with shape parameter approximately equal to two [1], and we therefore used two compartments for late-infection symptomatic individuals who will be hospitalised ($L_H$). For cases that recover without hospitalisation, in the absence of better data on the duration of infectivity since symptom onset, we made the parsimonious choice of a single late infection compartment with an exponentially distributed length of stay with mean 3.5 days, such that the overall period during which an individual is actively infectious (I plus the L compartments) is consistent with the 5-day mean generation time estimated in [24]. This level of accuracy is sufficient since we do not fit EpiBeds to non-hospitalised individuals. The resultant compartmental model is illustrated in Figure 3, with the state variables and parameters described in Tables 2a and 2b. The equations are reported in Section M.3.

**Table 2a:** *State variables for the compartmental model.*

| State variable | Description |
| --- | --- |
| S | Susceptible |
| $E_A, E_S$ | Exposed - will stay asymptomatic, become symptomatic |
| $I_A, I_S$ | Infectious - will stay asymptomatic, become symptomatic |
| $A_R$ | Asymptomatic - will recover |
| $L_R, L_H$ | Late infection (symptomatic) - will recover, be hospitalised |
| $H_R, H_C, H_D$ | Hospitalised - will recover, enter critical care, die without entering critical care |
| $C_M, C_D$ | Critical care - will be monitored before recovery, die |
| $M_R$ | Monitored - will recover |
| R | Recovered |

| D | Deceased |

**Table 2b:** *Parameter variables and values for the compartmental model. Where the chosen value is in Table A1, the chosen value is the one divided by the reported value.*

| Parameter variable | Description | Chosen value | Literature range | References |
|---|---|---|---|---|
| $r_E$ | Rate of transition through early stage infectious classes ($E_S$, $I_S$, $E_A$, $I_A$) | 1/4.85 | 1/4.85 | [1] |
| $r_{AR}$ | Rate of transition from late stage asymptomatic ($A_R$) to recovered (R) | 1/3.5 | (see text) | [24] |
| $r_{LR}$ | Rate of transition from late stage symptomatic ($L_R$) to recovered (R) | 1/3.5 | (see text) | [24] |
| $r_{LH}$ | Rate of transition from late stage severely symptomatic ($L_H$) | 1/5.2 | 1/5.2 | [1] |
| $r_{HR}$ | Rate of transition from hospital admission ($H_R$) to recovery (R), without ICU | Table A1 | 1/6.1 | [10] |
| $r_{HC}$ | Rate of transition from hospital admission ($H_C$) to ICU ($C_M$, $C_D$) | Table A1 | 1/4 to 1/1.5 | [8–10,19] |
| $r_{HD}$ | Rate of transition from hospital admissions ($H_D$) to death (D), without ICU | Table A1 | 1/9.8 to 1/7.5 | [25] |
| $r_{CM}$ | Rate of transition from critical care admission ($C_M$) to step down ($M_R$) | Table A1 | 1/16.8 to 1/12 | [10,14] |
| $r_{CD}$ | Rate of transition from critical care admission ($C_D$) to death (D) | Table A1 | 1/17 to 1/7 | [8,14,15] |
| $r_{MR}$ | Rate of transition from step down ($M_R$) to discharge (R) | Table A1 | 1/7 | [14] |
| $p_A$ | Proportion of infected individuals that will be asymptomatic | 0.55 | 0.179 to 0.972 | [26–30] |
| $p_H$ | Proportion of symptomatic individuals that will be hospitalised | 0.08 | 0.036 to 0.155 | [10,19,31] |
| $p_C$ | Proportion of hospitalised individuals that will enter critical care | Table 3 | 0.091 to 0.485 | [9,10,16,19,31] |
| $p_T$ | Proportion of hospitalised individuals that will die without entering critical care | Table 3 | 0.316 | [9] |
| $p_D$ | Proportion of individuals in critical care that will die | Table 3 | 0.4 to 0.453 | [9,14] |

We assumed only non-hospitalised infectious individuals contribute to new infections, with asymptomatic individuals less infectious than individuals who are pre-symptomatic or symptomatic. Due to behavioural changes, changes in test specificity, and the possibility that asymptomatic cases may correspond to individuals who simply have a long incubation period, identification of the relative infectivity of an asymptomatic case is challenging. We assume relative infectivity of 25%, based on [32,33]. We assume that asymptomatic cases make up 55% of infections, which we determined by adjusting age specific estimates of the asymptomatic rate to the age distribution in England [26]. Although infections from hospitalised patients could have an effect on the overall epidemic, most notably with health care workers as transmission links, detailed genetic data are required to characterise this process [34]. We also assume that nosocomial cases do not substantially alter hospital flow, i.e. upon testing positive nosocomial patients follow similar pathways to community-acquired cases. In the hospital admissions data, we either count patients from admission (if they were tested in the community) or from the date of their first positive swab result (if they were tested in hospital). This second cohort will include all nosocomial cases, who we treat as being admitted from the community.

## 2.3 Model fitting

### 2.3.1 Procedure

We fitted EpiBeds to English data (SITREP - NHS situation report and CPNS - COVID-19 Patient Notification System) using a Bayesian MCMC approach (full details on the fitting procedure are provided in Section M.4). To reduce the number of free parameters, we used the average waiting times in each hospital compartment estimated from the CHESS/SARI data (Table 1) and previously published estimates for disease parameters (Table 2b) as fixed model parameters. The remaining free parameters estimated from EpiBeds fit were: the initial number of infected individuals on 20/01/2020 (distributed between states $E_{A1}$ and $E_{S1}$ based on the proportion of asymptomatic infections, $p_A$), the transmission rates (which we assumed piecewise constant, with pre-selected change points as discussed in Section M.4.4, which generally correspond to large policy changes), the overdispersion of the Negative Binomial observation noise around each of the four data streams (one parameter per stream), and the probabilities of death if in ICU ($p_D$), of admission to ICU if hospitalised ($p_C$), and of death without going to ICU ($p_T$). For $p_D$ we used a strict Normal prior distribution with a mean and 95% CI for $p_D$ obtained from SARI data (Section M.4.3) for wave 1 of 35.7% (31.9%, 38.4%) and for wave 2 of 28.7% (26.5%, 32.1%). For the remaining free parameters, we used uninformative priors.

### 2.3.2 Results for the first and second wave

To demonstrate the performance of EpiBeds when fitting to data, we show results from fitting the first wave (01/03/2020 to 15/09/2020) and the second wave (01/08/2020 to 31/12/2020). We consider both waves independently to capture temporal changes in the hospital dynamics. We do not consider data from 01/01/2021, due to the large-scale vaccine rollout in England. Since there are substantial demographic shifts between the first and second wave, when fitting the second wave we used admissions for the whole time-series combined with beds, ICU, and deaths data only from 01/08/2020 onwards (see Section M.4.6). This enabled the probabilities to be fitted to the second wave independently of the first wave.

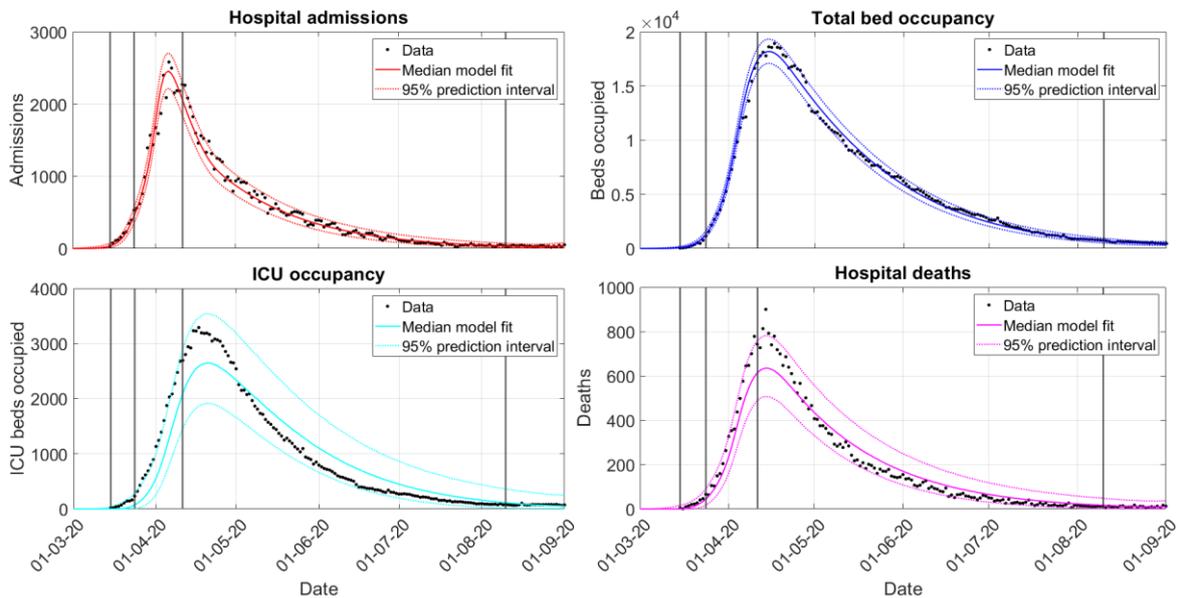

**Figure 4:** *Model performance for fitting the first wave to admissions (red), beds (blue), ICU beds (cyan) and deaths (magenta). Vertical lines indicate when transmission rate changes are added to EpiBeds (see Section M.4.4). Note that there is a delay between transmission changing and the hospital data streams changing, so inflection points occur after the transmission change point.*

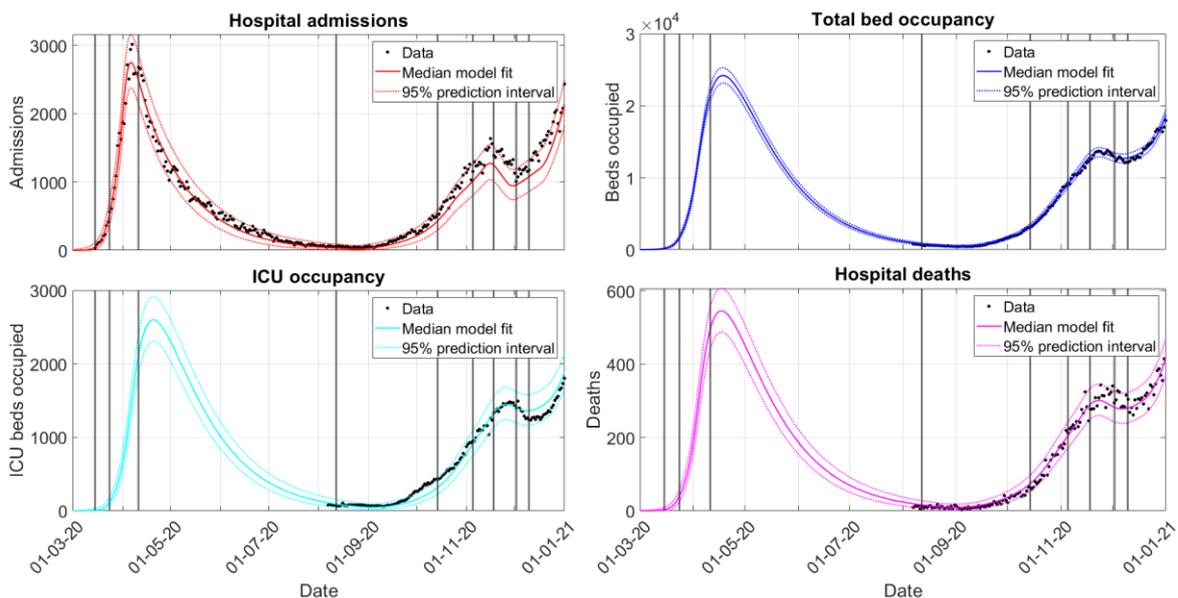

**Figure 5:** *Model performance for fitting the second wave to admissions (red), beds (blue), ICU beds (cyan) and deaths (magenta). Admissions data were fitted starting from 01/03/2020, while the other three data streams were fitted starting from 01/08/2020. Vertical lines indicate when transmission rate changes are added to EpiBeds (see Section M.4.4).*

Figures 4 and 5 show the output of the fitted model to the four hospital data streams for the first and second wave, respectively, demonstrating EpiBeds captures well the dynamics of both. The estimated overdispersion of data around the average model prediction shows that, for the first wave, EpiBeds fits admissions and hospital beds particularly well (low overdispersion), whereas ICU occupancy and deaths required high overdispersion to capture the data. This is driven by multiple factors including: (1) data quality issues between data streams at the start of the first wave (Section M.1), (2) a large shift in the age distribution of admissions from frailer older people in the spring to younger people with low mortality risk in the summer, and (3) changes in treatment which likely altered outcome probabilities.

For the second wave (Figure 5), there is better agreement among the data streams, due to more consistent reporting of data by the hospital trusts, less demographic shift in hospital admissions, and less dramatic changes in treatments, compared to the first wave. Although EpiBeds links all four data streams well during this period, there was a sharper increase in ICU admission during September 2020 than the model captured . During this period, admissions were concentrated in the relatively young, with severely ill younger patients more likely to visit ICU rather than be treated on the ward compared to older patients, since the former have more favourable ICU outcomes than the latter. As the epidemic spread through the community, the age distribution became relatively stable, corresponding to a slowdown in the ICU admission rate from October. Due to data quality issues in the early admissions data, we changed the data definitions used between the first and second wave slightly (see Section M.1), resulting in higher admissions in the data used when fitting the second wave. However, since for the second wave we only fitted the other three streams from 01/08/20 onwards (Section M.4.6), these data quality issues no longer affect the performance of EpiBeds at linking the four data streams.

Through fitting EpiBeds, we obtained two primary outcomes. First, we generated short term forecasts for each of the data streams by running the fitted model forwards in time. In Section 2.4, we investigate the accuracy at generating such forecasts. Second, we obtained posterior estimates for the free parameters (Section M.4.2). Some of these are model specific parameters, such as the overdispersion and initial conditions of the epidemic. However, others are of substantial interest in monitoring the epidemic. These are the probabilities of the different patient pathways and the transmission rates in the population (from which the effective and control reproduction numbers can be estimated). These posterior estimates change every time more data are added. For this reason, in Sections 2.3.3 and 2.3.4 we present only two representative sets of parameter estimates, for the first wave and the second wave. The reproduction number trajectory can retrospectively provide insight on the effectiveness of the interventions, and the differences in outcome probabilities can suggest changes, between the two waves, in the demographic of hospitalised patients, hospital policies and treatments, or a combination of these.

### 2.3.3 Outcome probability estimates

Outcome probabilities ($p_D$, $p_C$, $p_T$) were assumed to be constant throughout each wave and are presented only at the end of wave one (15/09/2020) and at the end of 2020 (31/12/2020), to highlight the difference between waves (Table 3). Using uninformative priors, EpiBeds did not provide reliable estimates for these three probabilities, though it still fitted the data well. Therefore, we used the CHESS/SARI data to determine an informative prior distribution for $p_D$. Whilst this is reasonable for $p_D$ due to the high coverage of ICU admissions in the CHESS/SARI datasets (over 90%), it was not possible to obtain similar priors for the probabilities $p_C$ and $p_T$ due to insufficient and geographically uneven coverage in the data, causing problems in both power and representativeness. We observe that, since the priors were quite strict on the values of $p_D$, the posterior estimates of $p_D$ generated through MCMC remained close to the prior, though a significant reduction is clearly visible. The estimated probability of being admitted to ICU ($p_C$), instead, has remained relatively constant throughout 2020. Finally, the probability of dying without ICU ($p_T$) has dropped by more than 25%, between the two waves. In Figure 6, we see that these posterior estimates are consistent with the range of estimates from the literature.

**Table 3:** *Posterior estimates for hospital pathway proportions. Brackets indicate 90% confidence intervals.*

| Parameter | Pre 15/09/2020 | Post 01/08/2020 |
| --- | --- | --- |
| $p_C$ | 0.125 (0.119, 0.130) | 0.127 (0.123, 0.129) |
| $p_T$ | 0.317 (0.305, 0.329) | 0.234 (0.229, 0.240) |
| $p_D$ | 0.344 (0.318, 0.372) | 0.296 (0.270, 0.321) |

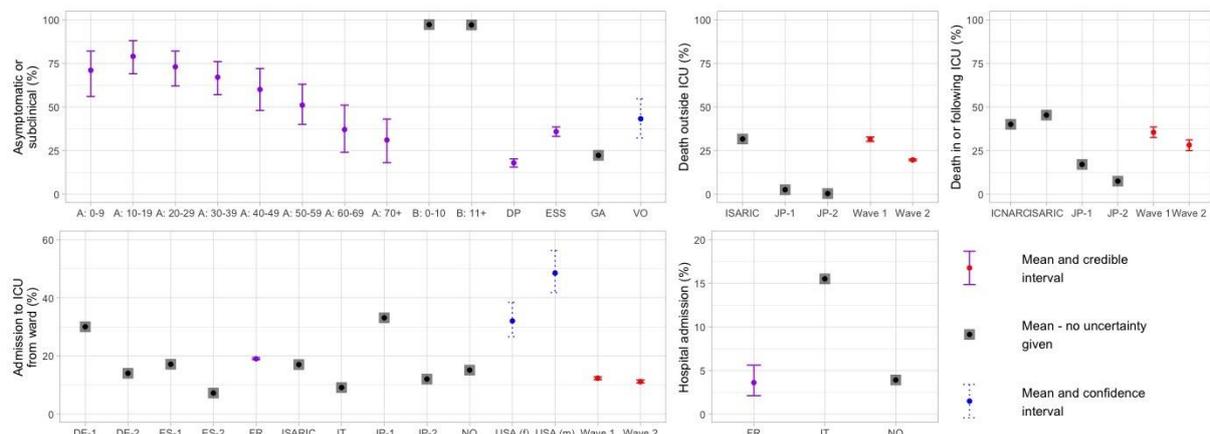

**Figure 6:** *Probabilities of asymptomatic individuals in the general population and different outcomes through the hospital pathway.* We present our own estimates of the probability of different hospital outcomes (red, labelled as Wave 1 and Wave 2), as well as estimates broken down across multiple age groups (A, [26]), estimates for Chinese children (B, [27]), and population averages in the German town of Gangelt (GA, [30]), on the Diamond Princess cruise ship (DP, [29]), across a Spanish serological survey (ESS, [35]), in a cohort study in California and Washington (USA, [16]) broken down by sex (m, f), and in the Italian municipality of Vo' (VO, [28]). Estimates are also obtained from ISARIC (ISARIC, [9]), ICNARC (ICNARC, [14]), France (FR, [19]), Italy (IT, [31]), Japanese hospital patients (JP, [17]), Spanish hospital patients in Reus (ES, [18]), German hospitalisations (DE, [36]), and Norway (NO, [10]). If not specified, all estimates are from the first wave. Otherwise, estimates are labelled for each wave but may not agree on the respective date ranges. The proportion of asymptomatic cases varies across studies and by age, whereas the proportion of individuals entering different hospital pathways is fairly consistent. Some of the presented quantities are obtained from estimating the proportions of patients with each outcome, and so no uncertainty is given. In other cases, these probabilities are inferred numerically, and so credible intervals are provided. There is clear geographic variability in hospital and ICU admission, which is known to correlate with age [31] and sex.

### 2.3.4 Reproduction number estimates

Using the transmission rates determined from EpiBeds, we estimated the reproduction number during the study period. Due to behavioural changes, implementation of control policies, and circulation of different variants, transmission rates have changed during the epidemic. Although there are multiple approaches to allow transmission changes, we opted for the parsimonious and tightly user-controlled one of only capturing major transmission changes, with dates determined by the introduction/relaxation of major interventions and/or visual inspection of the data (see Section M.4.4). Dates of major transmission changes used were:

- 13 March 2020 (visible change in hospitalisation trend, possibly due to media-driven behavioural changes or inaccuracies in recording early hospitalisation data),
- 24 March 2020 (beginning of a UK-wide lockdown),
- 11 April 2020 (visible change in trend towards the end of lockdown),

- 15 August 2020 (visible rise in hospital admissions),
- 6 September 2020 (visible change in trend),
- 14 October 2020 (Merseyside first area in England to enter "tier 3" restrictions),
- 5 November 2020 (England-wide second lockdown),
- 18 November 2020 (indicated by an increase in infections due to the rise of the B.1.1.7 variant -- now called Alpha -- in England and potentially increasing social interactions, this also encompasses any transmission changes after lifting the second lockdown on 2 December 2020).

In addition to these major transmission rate changes, whenever EpiBeds is fitted we include an additional change point 3 weeks before the final data point, unless a major intervention was already present within the last 3 weeks (see Section M.4.4). For the results presented here, this translates in additional transmission rate changes on:

- 10 August 2020, when producing the fit to the entire first wave (Figure 4),
- 10 December 2020, when producing the fit to the entire second wave (Figure 5).

Although further changes in transmission rates could have been added, this risked overfitting to noise in the data rather than genuine transmission trends.

We considered two reproduction numbers: the effective reproduction number and the control reproduction number [4]. The control reproduction number $R_c(t)$ is the average number of new infections generated by an average infection started at time $t$, in the absence of population immunity, assuming the transmission rate does not change over time (e.g. due to policy changes affecting physical distancing) from its value at time $t$. The basic reproduction number $R_0$ is then given by $R_c(t)$ before the first intervention reduces transmission by limiting the "natural" (i.e. pre-pandemic) population contact patterns. The effective reproduction number, $R_e(t)$ (also denoted $R_t$), describes the average number of new infections generated by an average infection started at time $t$, taking into account population immunity. This can be obtained by multiplying $R_c(t)$ by the susceptible fraction of the population at time $t$.

Since all susceptibles are assumed to be identical, and transmission rate constant in a given time window, the control reproduction number is given by the transmission rate multiplied by a constant factor, which encapsulates the probabilities of following each infection pathway, the relative transmission from each compartment and the average time spent in each compartment. Therefore, from the estimated transmission rates, we can calculate the control reproduction number $R_c(t)$ (Section M.3), and consequently the effective reproduction number $R_e(t)$ by scaling $R_c(t)$ by the proportion of remaining susceptible individuals (Table 4). Although $R_c(t)$ is constant throughout each interval, as the proportion of susceptibles changes continuously over time, so does $R_e(t)$. Therefore, we report the value of $R_e(t)$ only at the start of each interval in Table 4. Lockdown significantly reduced the transmission rate, and therefore $R_e(t)$. As lockdown went on, $R_e(t)$ increased slightly, as indicated by the transmission rate change on 19 April 2020. Over August, transmission increased, bringing $R_e(t)$ above 1. This growth continued until further interventions were brought in with the local tier system. This reduced the transmission rate, likely driven by the effectiveness of the tier 3 interventions in the North West. Finally, the second lockdown brought transmission down across the whole of England, bringing $R_e(t)$ below 1. Note that, using this model the initial reproduction number is not reliably constrained, since there are very few data points informing the initial transmission rate. This lack of identifiability is reflected in the MCMC trace plots (Figure M.1).

**Table 4:** *Posterior estimates for $R_e(t)$ and $R_c(t)$ values. Estimates marked with an asterisk are based only on admissions rather than all four data streams, so are potentially less reliable. Estimates marked with ^ are based on very few data points so will be unreliable.*

| Date of change | $R_e(t)$ - Pre 15/09/2020 | $R_c(t)$ - Pre 15/09/2020 | $R_e(t)$ - Post 01/08/2020 | $R_c(t)$ - Post 01/08/2020 |
| --- | --- | --- | --- | --- |
| **05/03/2020** | 5.87 (5.32, 6.54)^ | 5.87 (5.32, 6.54)^ | 5.44 (4.51, 6.35)*^ | 5.44 (4.51, 6.35)*^ |
| **15/03/2020** | 3.02 (2.91, 3.12) | 3.02 (2.91, 3.12) | 2.84 (2.64, 3.02)* | 2.84 (2.64, 3.02)* |
| **24/03/2020** | 0.67 (0.65, 0.68) | 0.67 (0.65, 0.68) | 0.78 (0.76, 0.81)* | 0.78 (0.76, 0.81)* |
| **11/04/2020** | 0.81 (0.80, 0.81) | 0.81 (0.80, 0.81) | 0.80 (0.79, 0.80)* | 0.80 (0.79, 0.80)* |
| **15/08/2020** | 1.54 (1.38, 1.71) | 1.61 (1.42, 1.79) | 1.62 (1.60, 1.65) | 1.71 (1.68, 1.74) |
| **25/08/2020** | 1.73 (1.51, 1.96) | 1.80 (1.58, 2.04) | NA | NA |
| **06/09/2020** | NA | NA | 1.48 (1.47, 1.49) | 1.57 (1.55, 1.68) |
| **14/10/2020** | NA | NA | 1.23 (1.22, 1.24) | 1.30 (1.29, 1.31) |
| **05/11/2020** | NA | NA | 0.80 (0.78, 0.83) | 0.85 (0.83, 0.88) |
| **18/11/2020** | NA | NA | 1.15 (1.13, 1.17) | 1.23 (1.21, 1.25) |
| **10/12/2020** | NA | NA | 1.43 (1.39, 1.46) | 1.54 (1.50, 1.57) |

We note that the longer the interval during which the transmission rate is assumed to be constant, the smaller the uncertainty. Moreover, the estimates of $R_e(t)$ that are obtained when only fitting the first wave are constrained by all four data streams, whilst the first wave $R_e(t)$ estimates obtained when fitting to the second wave are only constrained by the hospital admissions.

## 2.4 Short-term forecasting

To illustrate the performance of EpiBeds as a tool for real-time monitoring of the evolving pandemic, we present short-term projections made on days 1 and 15 of each month, from March to December 2020, based on the data available at that time, but on the background of the complete data. The posterior parameter estimates vary at every projection due to the additional data at each successive time point. We do not include here the specific parameter estimates, but only the projections for the data streams. The example results were generated using data that were available at the example time, to reflect how EpiBeds would have performed. See Section M.4.7 for details on the setup when generating these results.

The resultant forecasts are compared to the realised data streams in Figure 7. In the first forecasts (01/04/20), a transmission change was added on 24/03/20, to allow EpiBeds to adjust transmission based on lockdown. However, such a short fitting window leads to large uncertainty where the forecasts can either grow or decline (with the top of the y-axis truncated to aid visibility). By the 15/04/2020 forecast, the peak had been observed in the data. However, EpiBeds was unable to reconcile the four data streams, which resulted in the forecasts underestimating the reduction in the transmission rate and overshooting the data. This poor performance could be driven by multiple issues, such as challenges with estimating length of stay early in the pandemic (Section M.2, Appendix A), changing demographics after entering the first lockdown, and issues with the quality of the data streams (Section M.1). After this point, forecasts remained reliable into the summer. As transmission started to rise again, EpiBeds was able to accurately forecast the rise in all four data streams. However, throughout September and October, there was a demographic change, from younger to older age groups. This led to the ICU probability gradually declining and the mortality rate increasing, and the forecasts overestimated and underestimated, respectively, these two data streams.

In November, the demographic distribution of cases stabilised, and EpiBeds was able to reconcile all four data streams. At the end of the second lockdown the trend in the data was completely missed: this was to be expected, since EpiBeds generates forecasts by assuming the fitted transmission rates continue unaltered for the next two weeks, which was known not to be the case at the end of the lockdown.

Visually inspecting the figures, the 90% prediction intervals of the forecasts (shaded regions) appear to contain most data points. More precisely, 77% of data points, across all 4 data streams, fell within the prediction intervals. This varied across the data streams: admissions forecasts captured 76%; bed forecasts captured 80%; ICU forecasts captured 73%; and deaths forecasts captured 80%. Some of the cases when data points fall out of the 90% prediction interval occur where an intervention has been introduced within the forecasting window, causing the data to deviate from current trends. Others potentially arise from discrepancies between the data streams, particularly during the first peak, where data requests were first being picked up by the individual NHS trusts. Overall, this shows reasonably good model performance, and in practice throughout the pandemic EpiBeds has provided reliable forecasts in all regions considered. Note however that these percentages are only moderately informative in assessing model performance: for example, they would be much lower if the pandemic response had led to transmission rates changing very frequently, especially if the changes were de-synchronised with the dates the projections were made, since this would violate the assumptions of EpiBeds; conversely, during a long period of stable exponential growth or decline, model performance would be excellent.

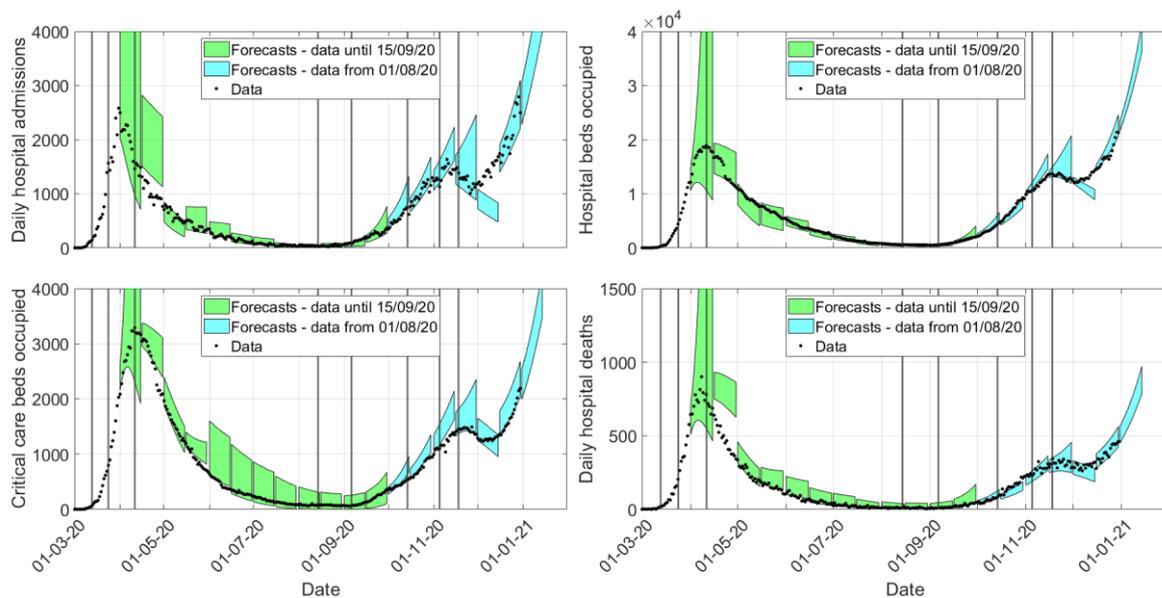

**Figure 7:** *England hospital forecasts. Green shaded regions are the 90% prediction intervals from forecasts up to 15/09/20. Blue shaded regions are the 90% prediction intervals from forecasts post 01/10/20 (using data from 01/08/20). Vertical black lines mark where major transmission changes occur, with changes in trajectory only manifesting after a delay that is data stream dependent. The y-axis is truncated to aid visibility, though a few forecast regions do exceed the y-limit.*

## 3. Discussion

To manage surges of COVID-19, hospital managers need accurate estimates of the length of stay of hospital patients, the proportion entering different hospital pathways, and an accurate tool for forecasting the number of beds likely to be occupied. We developed a compartmental model enabling short-term predictions for the flow of patients through hospitals. The length of stay in each hospital compartment was estimated using detailed line-list data, with the results used to parameterise EpiBeds. The novelty of this model compared to standard compartmental models is the explicit inclusion of compartments depending on patient outcomes, thereby enabling optimal use of available data whilst keeping model complexity low. EpiBeds was fitted to occupancy data enabling us to estimate the proportion of patients entering each hospital pathway, generate short-term hospital occupancy predictions, and help inform management of hospital caseloads. EpiBeds further enabled us to estimate the effective and control reproduction numbers during different periods of the epidemic, corresponding to substantial changes in the hospital trends driven by major policy changes, the emergence of new variants, and seasonal effects.

Here, we fitted EpiBeds to hospital data for England, but it can readily be applied to other geographies. As part of the COVID-19 response, it was used to generate forecasts for Scotland, Wales, Northern Ireland, and the United Kingdom, as well as for smaller English regions. In line with other published estimates [3,19,31], we estimated 12% of COVID-19 patients were admitted to ICU, during both the first and second waves in England. The proportion of patients surviving on ICU improved over time, with 34% mortality during the first wave and 30% during the second wave. An even stronger reduction in mortality occurred outside the ICU, with 32% of admissions dying without ICU during the first wave and 23% during the second wave. This reflects the change in the age distribution of cases and potential improvements in treatments. Given only 12% of admitted patients went to ICU, the vast majority of deaths occurred outside of ICU: 1/3 and 1/4 of all admissions outside ICU, in the first and second wave, against 4% or less in ICU – i.e. about 89% and 84% of all deaths in the two waves, respectively. In most cases these were frail individuals for whom ICU was unsuitable.

We validated the short-term forecasting performance of EpiBeds by generating 14-day forecasts using data at the start and midpoint of each month. Most of the forecasts captured the data within the 90% prediction interval of the forecasts, demonstrating the reliability of EpiBeds for providing short-term hospital flow predictions. When transmission rates are stable, forecasting accuracy is very high. However, if transmission rates change, driven for example by major policy changes, forecasts can deviate from the data. Data quality issues can also affect predictions, likely driving some of the forecasting inaccuracies observed during the first wave. Understanding these potential data issues is important since otherwise the divergence from the data may be attributed to model misspecification. The minimal complexity of EpiBeds makes it easy to identify what causes issues in the model fitting and makes both model behaviour and model limitations transparent, which we deem to be one of the key reasons to advocate for the use of simple models.

Our compartmental model was developed specifically to provide predictions of hospital occupancy and designed to maximise the information in available data whilst minimising the inclusion of unsupported assumptions. For this reason, EpiBeds is not structured by sex or age, or other comorbidities such as heart failure or chronic kidney disease, even though these are known to affect disease severity [13,37]. The SITREP does not include sex as a category and does not include age for all data streams (particularly when the model was first developed). As epidemics progress, the communities in which the virus circulates may change, which in turn could affect how individuals progress through hospital pathways, such as the probability of entering ICU if critically ill. This emphasises the need for the consistent reporting of high-quality data so that estimates can be continuously updated, resulting in more accurate forecasts. To account for demographic changes, as well as potential improvements in treatments, we fitted the parameters for the second wave independently of the first wave.

We have presented a minimally complex compartmental model of COVID-19 hospital pathways, specifically tailored to available data sources. Model predictions were accurate, except when transmission rates changed dramatically, and enabled retrospective estimates of reproductive values. Our model differs from more conventional compartmental models by defining compartments based not only on current status, but on future outcome, making it more closely aligned to the data. The relative simplicity of EpiBeds means it can be used to make short-term predictions in different settings, as well as used as a framework to develop short-term forecasts in the case of new outbreaks.

# Methods

*M.1 Data*

Several datasets were used: CHESS (COVID-19 Hospitalisations in England Surveillance System), which has since been replaced by SARI (Severe Acute Respiratory Illness); the SITREP (NHS situation report); and the CPNS (COVID-19 Patient Notification System).

*M.1.1 CHESS/SARI*

CHESS/SARI records patient-level data detailing the date of each event in their hospital pathway, such as time of hospital admission and discharge, time of ICU admission and discharge, and time of death in hospital. This provides detail into how long individuals spend in hospital and the type of pathways they take.

However, this is not a complete data set, as it only includes a subset of COVID-19 positive hospital patients that are chosen for submission by the individual hospital trusts. This sample is heavily biased towards patients that require critical care, since some hospital trusts only include critical care patients in their submissions, whereas others include all patients. Due to this, CHESS/SARI are not reliable indicators of the overall hospital burden.

Using these data, we primarily estimated the conditional length of stay distributions. Whilst incomplete, this data is suitable for estimating length of stay conditioned on specific outcomes, provided these conditional samples are representative of the wider population. With the large sample size of the CHESS/SARI data, this should be a reasonable assumption. In addition to length of stay, we also used these data to estimate a prior distribution for the probability that patients die in ICU, since ICU patient coverage is very high.

*M.1.2 SITREP*

SITREP contains the number of COVID-19 positive patients in various areas of the hospital. These data includes the number of admissions each day who had a positive test before admission, the daily number of inpatients who have been newly diagnosed with COVID-19, the number of COVID-19 patients in any hospital bed and the number of COVID-19 patients in ICU beds. Unlike the CHESS data, this is a mostly complete data set, though may be subject to occasional missingness, since it relies on daily submissions from individuals NHS trusts, with no historic corrections if days are missed.

In the early days of the pandemic, data quality issues led to potential inconsistencies in the data submitted between different hospital trusts. The SITREP admissions data comes in two streams: "patients tested before admission" and "patients tested after admission". In the first wave, most patients were tested after admission, since testing rates in the community were low. To adjust for the data quality issues, using only the "tested after admission" data stream, rather than summing these together, worked better for the first wave. By the second wave, these early data quality issues had been solved, so we summed the two data streams together when fitting admissions data. This resulted in the modelling prior to 15/09/2020 having fewer hospital admissions during the first wave than the post 01/08/2020 modelling. Despite the potential inconsistencies in these data, the SITREP provides the most reliable time-series for hospitalisation data in England.

Using the SITREP data, we constructed three time-series: daily hospital admissions, daily hospital bed occupancy, and daily critical care (ICU) bed occupancy. Note that the hospital bed occupancy includes both critical care and non-critical care beds. These provided three of the four data streams used to fit EpiBeds.

*M.1.3 CPNS*

CPNS is a complete line list of all COVID-19 positive individuals who have died in English hospitals. From this line-list, we can construct a time series of the number of COVID-19 hospital deaths each day. This should be a complete set of all COVID-19 deaths in hospital.

Note that, although this records all COVID-19 flagged deaths in hospital, these may not be exclusively caused by the disease. These use the definition created by PHE of deaths within 28 days of a positive test. Some of these may be attributed to other underlying causes. However, these figures are very close to the number of deaths with COVID-19 listed as a cause of death, with 100,000 reported by the ONS [38] compared to 95000 by PHE [39], as of 18/01/2021.

The CPNS data were used to construct a time-series for daily hospital deaths. Combining this with the three SITREP-derived data streams gave the four data streams used to fit EpiBeds.

*M.2 Estimating delay distributions*

Delay distributions describe the distribution of times between subsequent events. In the disease progression pathway, there are many such distributions, including the incubation period (i.e., the length of time from infection to symptom onset), the serial interval (i.e., the time from symptom onset of an infector to symptom onset of their infectee), and lengths of stay in different hospital compartments.

Although the data contains information on these delays, we typically do not directly observe the delay but instead the timing of the two events. Therefore, this information needs to be considered when fitting the data. We are interested in finding a delay distribution that maximises the likelihood of the observed event dates. To construct a likelihood function, we condition on the time of the first event and look forwards to the probability of the second event occurring on the observed date. However, since we are collecting data during an ongoing outbreak, the sample is truncated. This means that for cases admitted recently, the second event will only be observed if the delay is short, which will overexpress short delays. To account for this, we condition the likelihood function against the second event being observed before the truncation date [40].

Let $E_1$ denote the first event and $E_2$ denote the second event. The truncation date will be denoted by $T$. Therefore, for a sample $x$ with $E_1 = e_1$ and $E_2 = e_2$, the corresponding likelihood function is [40]

$$P(E_2 = e_2 | E_1 = e_1 \cap E_2 \leq T) = \frac{P(E_2 = e_2 \cap E_1 = e_1 \cap E_2 \leq T)}{P(E_1 = e_1 \cap E_2 \leq T)}$$
$$= \frac{P(E_1 = e_1)P(E_2 = e_2 | E_1 = e_1)}{P(E_1 = e_1)P(E_2 \leq T | E_1 = e_1)}$$
$$= \frac{P(E_2 - E_1 = e_2 - e_1)}{P(E_2 - E_1 \leq T - e_1)} = \frac{f_\theta(e_2 - e_1)}{F_\theta(T - e_1)}$$

where $f_\theta(.)$ is the probability density function and $F_\theta(.)$ is the cumulative density function of the delay between events $E_1$ and $E_2$, parameterised by $\theta$.

To fit the distributions, we need to process the data into the required format. Since this likelihood function conditions on both events being observed before the truncation date, we condition the sample for each delay to only include cases for which both events have been observed. This removes censoring issues present in the CHESS data, and the truncation correction should correct for the fewer long delays observed. When fitting the delays, we fit the first and second waves independently, to account for changes in these delays. For the first wave, we use data from 01/03/2020. We generate estimates using data available up to the end of each month, until 31/08/2020. For the second wave, we use data from 01/08/2020, again generating estimates using data available up to the end of each month, until 31/12/2020.

To parameterise EpiBeds, for simplicity we only use the maximum likelihood estimates for the mean and the standard deviation each of these delays, rather than the more comprehensive approach of using prior distributions based on uncertainty around the point estimates. More precisely, we find the integer closest to the estimated shape parameter and use that to inform the model structure by placing the same number of identical compartments in series ("linear chain trickery"). We then use the point estimate for the mean to compute the transition rates.

However, to provide information of the reliability of the point estimates, we still report uncertainty around them, generated through parametric bootstrapping. Assuming that our data are a sample from the chosen family of distribution (in this case Gamma) parametric bootstrapping requires generating a random sample from the Gamma distribution parameterised with the point estimates, with the same sample size as the original data. By fitting another Gamma distribution to this sample, we can obtain new point estimates. Repeating this numerous times allows us to construct uncertainty around our point estimates from the original model fitting. For the first and second waves, the overall length of stay distributions, with uncertainty from parametric bootstrapping with 1,000 repeats, are presented in Table 1. Monthly point estimates of the mean delay, used as input for the short-term forecasts in Section 2.4, are given in Table A1.

*M.3 The ODE compartmental model for hospital flow*

A deterministic framework is appropriate to model the flow between hospital compartments, since hospital capacity is of greatest concern when case numbers are high, and hence stochastic effects are limited. The model structure obtained in Section 2.2 describes a compartmental model between different states in the hospitalisation pathway, coupled with a background epidemic model. To model the flow between states, we used a system of ordinary differential equations. We accommodated noise in the data by assuming they were generated from a Negative Binomial distribution about the solution to the ordinary differential equations. This enabled us to construct a likelihood for the data being generated by EpiBeds (see Section M.4.5).

In an ODE compartmental model with constant progression rates, the permanence waiting times in each compartment are exponentially distributed with mean equal to the inverse of the rate. By a slight modification of the model, one can describe Erlang distributions of permanence times: an Erlang distribution with shape parameter $n$ corresponds to the sum of $n$ independent and identically distributed exponential distributions. Additional "hidden" compartments can therefore be added, between which the delays will be exponentially distributed [41]. In particular, if all rates are equal to $r$, the mean permanence time in a sequence of $n$ compartments will be Erlang distributed with mean $n/r$ and shape parameter $n$ [22]. This way of representing Gamma-distributed permanence time is also known as "linear chain trickery".

We used the linear chain trickery to construct a model where the delay distributions between hospital compartments are good approximations of the delay distributions estimated from data (see Sections 2.1 and M.2). In particular, if the mean and standard deviation of a delay distribution are similar, then the resulting Gamma distribution is very close to being exponentially distributed, meaning the source population can be described by a single compartment. If the standard deviation is smaller than the mean, then the Gamma distribution may be closer to an Erlang distribution with shape parameter greater than 1.

Since all hospital length of stay distributions are approximately exponential (because the mean is similar to the standard deviation), we model the hospital compartments using constant rates. For the background epidemic, we represented the Gamma-distributed incubation period (shape parameter 3, [42]) and the Gamma-distributed time between symptom onset to hospitalisation (shape parameter 2,

[1]), by using three and two compartments in sequence. This results in EpiBeds described by the flowchart in Figure 3 and by the following system of ODE's (where the time index in all parameters and variables has been dropped for brevity)

$$\frac{dS}{dt} = -\lambda$$

$$\frac{dE_{S_1}}{dt} = (1-p_A)\lambda - 3r_E E_{S_1}$$

$$\frac{dE_{S_2}}{dt} = 3r_E E_{S_1} - 3r_E E_{S_2}$$

$$\frac{dE_{A_1}}{dt} = p_A \lambda - 3r_E E_{A_1}$$

$$\frac{dE_{A_2}}{dt} = 3r_E E_{A_1} - 3r_E E_{A_2}$$

$$\frac{dI_S}{dt} = 3r_E E_{S_2} - 3r_E I_S$$

$$\frac{dI_A}{dt} = 3r_E E_{A_2} - 3r_E I_A$$

$$\frac{dL_R}{dt} = 3r_E I_S (1-p_H) - r_{LR} L_R$$

$$\frac{dA_R}{dt} = 3r_E I_A - r_{AR} A_R$$

$$\frac{dL_{H_1}}{dt} = 3r_E I_S p_H - 2r_{LH} L_{H_1}$$

$$\frac{dL_{H_2}}{dt} = 2r_{LH} L_{H_1} - 2r_{LH} L_{H_2}$$

$$\frac{dH_D}{dt} = 2r_{LH} L_{H_2} p_T - r_{HD} H_D$$

$$\frac{dH_C}{dt} = 2r_{LH} L_{H_2} p_C - r_{HC} H_C$$

$$\frac{dH_R}{dt} = 2r_{LH} L_{H_2} (1 - p_T - p_C) - r_{HR} H_R$$

$$\frac{dC_D}{dt} = r_{HC} H_C p_D - r_{CD} C_D$$

$$\frac{dC_M}{dt} = r_{HC} H_C (1 - p_D) - r_{CM} C_M$$

$$\frac{dM_R}{dt} = r_{CM} C_M - r_{MR} M_R$$

$$\frac{dD}{dt} = r_{HD} H_D + r_{CD} C_D$$

$$\frac{dN}{dt} = -r_{HD} H_D - r_{CD} C_D$$

$$\frac{dR}{dt} = r_{HR} H_R + r_{MR} M_R + r_{LR} L_R + r_{AR} A_R,$$

where

$$\lambda = S\beta\big(f(I_A + A_R) + I_S + L_R + L_{H_1} + L_{H_2}\big).$$

Here, $f$ is the reduction in transmission for asymptomatic cases, which is taken to be $f = 0.25$.

From the solution to the ordinary differential equations, the control and effective reproduction numbers can be calculated. The control reproduction number is given by

$$R_c(t) = \beta(t)k,$$

where

$$k = (1 - p_A)\left(\frac{1}{3r_E} + \frac{p_H}{r_{LH}} + \frac{1 - p_H}{r_{LR}}\right) + p_A f\left(\frac{1}{3r_E} + \frac{1}{r_{AR}}\right).$$

From the control reproduction number, the effective reproduction number can be calculated as

$$R_e(t) = R_c(t)\frac{S(t)}{N(t)}.$$

## M.4 Model fitting procedure

In this section, we provide details on all the steps involved in the fitting procedure, as briefly outlined in Section 2.3.1. First, we describe the general model fitting procedure. We then describe how EpiBeds is fitted to different waves of the epidemic, before describing how EpiBeds was set up when retrospectively generating the forecasts to investigate their accuracy. These reflect how EpiBeds would have been run at the time, showing how to set it up when generating new forecasts.

### M.4.1 Fixed parameters

In EpiBeds, we assume that certain parameters are fixed (Table M.1), taking points estimates from the literature or our own methods (Section M.2). We choose to fix these estimates since we are confident in their reliability, and having fewer free parameters reduces potential identifiability issues when fitting the model. EpiBeds could be extended to use prior distributions for these parameters, which may also aid identifiability, but in its current form we opt to use fixed parameters. This decision may reduce some of the uncertainty in the projections, but still results in reliable fits and forecasts.

**Table M.1:** *List of fixed parameters for the compartmental model*

| Variable | Description |
| --- | --- |
| $r_E$ | Rate of transition from early stage infectious classes ($E_S$, $I_S$, $E_A$, $I_A$) |
| $r_{AR}$ | Rate of transition from late stage asymptomatic ($A_R$) to recovered (R) |
| $r_{LR}$ | Rate of transition from late stage symptomatic ($L_R$) to recovered (R) |
| $r_{LH}$ | Rate of transition from late stage severely symptomatic ($L_H$) |
| $r_{HR}$ | Rate of transition from hospital admission ($H_R$) to recovery (R), without ICU |

| $r_{HC}$ | Rate of transition from hospital admission ($H_C$) to ICU ($C_M$, $C_D$) |
| --- | --- |
| $r_{HD}$ | Rate of transition from hospital admissions ($H_D$) to death (D), without ICU |
| $r_{CM}$ | Rate of transition from critical care admission ($C_M$) to step down ($M_R$) |
| $r_{CD}$ | Rate of transition from critical care admission ($C_D$) to death (D) |
| $r_{MR}$ | Rate of transition from being stepped down ($M_R$) to discharge (R) |
| $p_A$ | Proportion of infected individuals that will be asymptomatic |
| $p_H$ | Proportion of symptomatic individuals that will be hospitalised |

*M.4.2 Free parameters*

When fitting EpiBeds, we have multiple free parameters that need to be estimated (Table M.2). These pertain to parameters that cannot be directly inferred from high quality data and parameters describing the background epidemic, which is not directly observed and therefore must be inferred. By fitting EpiBeds to the data, we obtain estimates for these crucial parameters, which can describe the risk posed to patients in hospital and the reproduction number of the background epidemic.

**Table M.2:** *List of free parameters for the compartmental model*

| Variable | Description |
| --- | --- |
| $I_0$ | Initial number of infected individuals (divided in proportions given by $p_A$ and $1-p_A$ between the first $E_A$ and $E_S$ stages) |
| $\sigma_A$ | Overdispersion parameter for admissions data |
| $\sigma_B$ | Overdispersion parameter for beds data |
| $\sigma_C$ | Overdispersion parameter for ICU data |
| $\sigma_D$ | Overdispersion parameter for deaths data |
| $p_C$ | Proportion of hospitalised individuals that will enter critical care |
| $p_T$ | Proportion of hospitalised individuals that will die without entering critical care |
| $p_D$ | Proportion of individuals in critical care that will die |
| $\beta_1$ | Transmission rate up to 15/03/2020 |
| $\beta_2$ | Transmission rate from 15/03/2020 to 24/03/2020 |
| $\beta_3$ | Transmission rate from 24/03/2020 to 11/04/2020 |
| $\beta_4$ | Transmission rate from 11/04/2020 to 15/08/2020 |
| $\beta_5$ | Transmission rate from 15/08/2020 to 06/09/2020 |
| $\beta_6$ | Transmission rate from 06/09/2020 to 14/10/2020 |

| $\beta_7$ | Transmission rate from 14/10/2020 to 05/11/2020 |
| $\beta_8$ | Transmission rate from 05/11/2020 to 18/11/2020 |
| $\beta_9$ | Transmission rate from 18/11/2020 to 10/12/2020 |
| $\beta_{10}$ | Transmission rate from 10/12/2020 to 31/12/2020 |

*M.4.3 Prior estimates for outcome probabilities*

To generate prior estimates for the probability of dying on ICU we used the CHESS/SARI dataset, which contains approximately 90% of the English ICU data (based on comparing patient counts to the SITREP data). This suggests SARI ICU outcomes are representative of general ICU outcomes.

Due to the censoring of patient outcomes, and to capture data uncertainty, we estimate outcome probabilities using a non-Markovian competing risk model [25]. By fitting this model to the data, we can estimate the hazard functions for these competing risks. From these fitted hazard functions, we can simulate posterior distributions for the outcome probabilities, which allows us to quantify uncertainty around the point estimates. In the original paper, the authors describe a bootstrapping method for simulating uncertainty. However, a more recent version of the model, using MCMC to generate uncertainty, has been described on the corresponding Github [43], which we use here. We consider the outcome probabilities before and after September 2020, to model the wave 1 and wave 2 outcomes, with the results summarised in Table 3. We also report the estimates for $p_D$ obtained at the end of each month since 01/03/2020 (Table A.2), which gives the prior estimates used to generate the short-term forecasting scenarios in Section 2.4.

*M.4.4 Transmission rate changes*

In EpiBeds, the background epidemic is driven by a transmission rate $\beta$, that represents the total infectious pressure exerted by a symptomatic infectious individual in their early and late infectious stages. For the asymptomatic infectious classes, we assume the infectious pressure exerted is a fraction of this value. This parameter collates contact behaviour, transmission probability of contacts and strength of contacts into a single parameter. On an individual level, this does not provide accurate information about the transmission dynamics, but on a population level aggregating all of these into a single parameter is a simple way to represent the average transmission dynamics in the population.

To model the background epidemic, we need to estimate the value of this transmission parameter. We cannot assume this is constant, because there are large changes in this parameter as behaviour changes, for example due to lockdown. On the other hand, we do not want to add too many different values, as this risks overfitting noise in the data rather than genuine changes in transmission. There are various clear transmission rates that need to be fitted: the transmission rate that led to the original growth of the outbreak, the transmission rate after lockdown 1, the transmission rate when lockdown 1 is eased, the transmission rate after lockdown 2, and the transmission rate after lockdown 2 is eased. The easing of lockdown 1 was gradual, so it is not easy to specify dates for the transmission rate changes. Through looking at the data on an exponential scale, we notice two changes in the gradient that can be associated with the easing of lockdown 1. Since this data is on hospitalisations, we consider these gradient changes to correspond to transmission changes occurring 9 days prior [1], which corresponds to the median delay between infection and hospital admission. Therefore, we add new transmission rates on 11/04/2020 and 15/08/2020. From the log-scale, we also notice that transmission changed before the first lockdown. This was not sufficient to bring cases down, but slowed down the original rate of growth. Therefore, we add a new transmission rate on 13/03/2020. Transmission trends changed as the second wave became established, so a further change point was added on 06/09/2021. Additionally, although not a national policy, strict restrictions were imposed in the North West in October, before the second national lockdown in November. Since at this time

North West had the highest prevalence in England, this led to a substantial change in the national trends. Therefore, we add another transmission change on 14/10/20 to correspond to the "tier 3" restrictions imposed in Liverpool. Although the second national lockdown was lifted on 02/12/2020, it became apparent that transmission had already increased before this date, likely driven by a combination of increased movement towards the end of lockdown and the emergence of the more transmissible B.1.1.7 variant [44]. To capture this, investigating the data on a log-scale suggests a transmission change should be added on 18/11/2020. This transmission change also encompasses any transmission change from lifting the second lockdown, so we do not explicitly add the 02/12/2020 change point.

In theory, interventions can be added to EpiBeds as soon as they have occurred. However, with a median delay of 9 days from infection to hospital admission, there may be insufficient data to inform EpiBeds on the impact of the intervention immediately after their implementation. To remove unnecessary uncertainty, if an intervention occurs within the last 7 days of data, we do not include this, so that the forecast will instead generate a counterfactual where the intervention has no impact on transmission. On the other hand, significant transmission changes that do not correspond to known intervention dates cannot be added until there is sufficient data to observe the change in trend. In practice, we only consider such changes once half a month has passed, by which point there will be sufficient data to quantify the change.

In addition to these fixed transmission changes, when generating forecasts we consider a final change point 3 weeks prior to the final data point, unless there is another change point in this interval. This is to allow EpiBeds to react to changes in transmission to predict a current value of transmission, though this will be a transmission averaged over the last 3 weeks rather than a now-cast. A 3-week window was chosen as a bespoke compromise between an interval that is too long, which would give narrow uncertainty and estimates of $R_e(t)$ excessively slow at responding to visible changes in the data trends, and one that is too short to ensure there is sufficient data to inform the final transmission rate (recall that, with an average delay of 9 days from infection to hospitalisation, the last few data points provide less and less information on transmission as they approach the time at which projections are made) and estimates would be dominated by random noise and hard-to-control discrepancies between the signal from different data streams. When this 3-week transmission change suggests a substantial change to the previous trend, this prompts investigations into whether a new fixed transmission change should be added, following the procedure described above.

Through this, we obtain a system of ordinary differential equations with a time varying parameter beta, which changes values according to a step function. The jump times of this step function are pre-specified, but the resulting values are fitted to the data.

*M.4.5 MCMC*

Adding Negative Binomial noise to the ODEs (Section M.3) enables us to calculate a likelihood function for observing the data given our model parameters. This is based on the probability that the deviation between our model and the data can be explained by noise. For each of the four data streams, we can construct a likelihood function for that data, which can then be multiplied together to build the overall likelihood function. To improve computational stability, we consider the log-likelihood function:

$$\begin{aligned} L = &\sum \ln(f(d_A, y_A/(\sigma_A - 1), 1/\sigma_A)) \\ &+ \sum \ln(f(d_B, y_B/(\sigma_B - 1), 1/\sigma_B)) \\ &+ \sum \ln(f(d_C, y_C/(\sigma_C - 1), 1/\sigma_C)) \\ &+ \sum \ln(f(d_D, y_D/(\sigma_D - 1), 1/\sigma_D)) \end{aligned}$$

Where A, B, C and D refer to the four different data streams fitted (see e.g. Table M.2) and, for each of them, σ is the overdispersion parameter of the Negative Binomial observation noise, d is the data and y is the solution to the ODEs. The continuous variables y are defined as

$$y_A = r_{LH} L_{H_2}$$
$$y_B = H_D + H_C + H_R + C_D + C_M + M_R$$
$$y_C = C_D + C_M$$
$$y_D = r_{HD} H_D + r_{CD} C_M,$$

and are evaluated at each day for which a data point d is available. The sums are over all days for which data is available.

Using this likelihood function, we can estimate the parameters of EpiBeds. We develop an MCMC algorithm, which allows us to navigate the challenging parameter space to find our optimal solution, and generate reliable uncertainty bounds around the solution. There are multiple unknown parameters that we need to fit (Section M.4.2, Table M.2). For the probability of dying on ICU, we need to use an informative prior (Section M.4.3), which we implement through adding the prior to the likelihood function, giving:

$$\begin{aligned} L = &\sum \ln(f(d_A, y_A/(\sigma_A - 1), 1/\sigma_A)) \\ &+ \sum \ln(f(d_B, y_B/(\sigma_B - 1), 1/\sigma_B)) \\ &+ \sum \ln(f(d_C, y_C/(\sigma_C - 1), 1/\sigma_C)) \\ &+ \sum \ln(f(d_D, y_D/(\sigma_D - 1), 1/\sigma_D)) \\ &- (\frac{1}{2}\ln(2\pi\sigma_{prior}^2)) - \frac{1}{(2\sigma_{prior}^2)(p_D - \mu_{prior})^2)} \end{aligned}$$

where $\mu_{prior}$ is the mean prior estimate of $p_D$ and $\sigma_{prior}$ is the standard deviation of the prior $p_D$ estimate.

To fit EpiBeds, we use a manually tuned random walk MCMC algorithm implemented in Julia. The input data depends on whether the first wave or second wave is being fitted, as described above. We start the epidemic 40 days before the first data point, a bespoke value which should be large enough to allow the number $I_0$ of initial cases, which only occupy to the first $E_A$ and $E_S$ states, to flow into all other compartments and occupy them in roughly stable proportions (which depend on the initial rate of growth, and hence on the first fitted transmission rate $\beta_1$), before the model solution reaches the first data point. Therefore, the first day considered in EpiBeds is 20/01/2020, with the first data point on 01/03/2020. Prior values for EpiBeds parameters are specified as described in Sections M.4.1 – M.4.3, coupled with initial conditions for the free parameters with uninformative priors. The ODE is then solved for the input parameters, generating the time-series output that are added to the likelihood functions. Based on these likelihoods, the parameter values are scored and resampled, allowing EpiBeds to explore the parameter space. Code for simulating EpiBeds, and generating the scenarios shown in the paper, are available at [45], along with trace plots for all MCMC results included in this paper. Unfortunately, input data cannot be shared, since this was provided through a data sharing agreement, but similar publicly available data are available at [39].

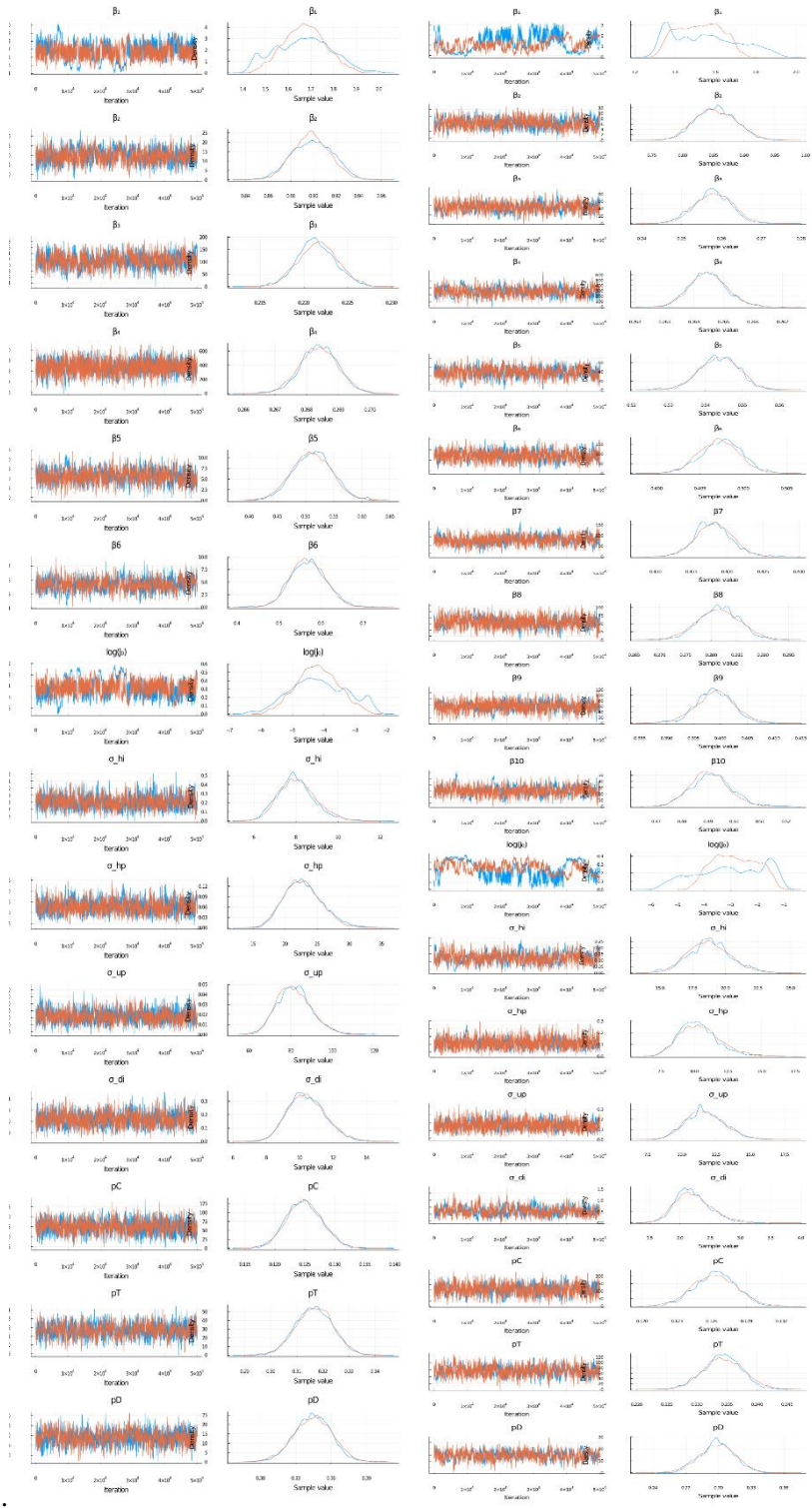

**Figure M.1.** *MCMC trace plots for EpiBeds for the results shown in Figures 4 and 5. Note that $\beta_1$ and $J_0$ are highly correlated and poorly identified.*

### M.4.6 Fitting different waves

When fitting the data, we consider the first and second waves in a different manner. Due to changes in length of stay and patient outcomes over time, we cannot fit a single set of parameters over the whole pandemic. Therefore, we fit the two waves independently. For the first wave, we include the full data for all data streams from 01/03/2020 until 15/09/2020. For the second wave, we include the full admissions data from 01/03/2020 until 31/12/2020, which is used to constrain the background epidemic, and for the other three streams we only use data from 01/08/2020 until 31/12/2020. This enables different length of stay estimates to be used for each wave and allows us to estimate different outcome probabilities for each wave.

First wave

To fit the first wave of the epidemic, we use all four data streams, using data starting on 01/03/2020. We start the simulations from 40 days prior to this, starting on day 20/01/2020. This is to allow the initial conditions to treacle through all compartments before the first data point on 01/03/2020. During this initial phase, EpiBeds fits an initial number of infected and initial transmission rates that allow the model to meet the data. EpiBeds is then constrained by all four data streams, determining the magnitude of the pre-specified transmission rate changes and the relationships between the data streams.

Second wave

When fitting the second wave, we remove beds, ICU, and deaths data prior to 01/08/2020. Prior to this date, EpiBeds is only constrained by the hospital admissions data, and only the first term of the likelihood (which does not depend on the outcome probabilities $p_C$, $p_T$ and $p_D$) is used. As above, we start EpiBeds on 20/01/2020 in order to meet the hospital admissions data on 01/03/2020. Post 01/08/2020, we introduce the other three data streams and compute the other likelihood terms. This then constrains the probabilities to fit the relationship between these data streams in the second wave.

*M.4.7 Evaluating the accuracy of short-term forecasting*

Since this model was developed for short-term forecasts for national planning, it is important to analyse model performance. Throughout the pandemic, EpiBeds has been continuously refined. Therefore, instead of analysing the historic forecasts that were generated, we generate new forecasts retrospectively, using data that were available at the time. We generate two forecasts per month, one on the first day of each month and one on the 15th day, generating 14-day forecasts to ensure the forecasts are non-overlapping. This is to aid visualisation whilst covering all phases of the epidemic.

Data used

When generating forecasts, we use SITREP and CPNS data up to the day before the first day of forecasts. In this data, hospital admissions have the last two days removed and deaths have the last five days removed, since these data points are lagged and will be revised upwards. For the length of stay estimates and prior $p_D$ estimate, we use data up to the end of each month. Therefore, for forecasts generated on the 1st, the data run up to the previous day, whereas for the 15th the data run up to 15 days before. It is possible to continuously generate these estimates in real-time, but this is not reflective of the workflow in practice. Early in the epidemic, this may have had an influence, since length of stay estimates were changing quickly due to right-censoring and truncation in the data. However, after a couple of months the estimates became more stable, so this choice will not have a substantial influence on the results.

Transmission rate changes

As described in Section M.4.4, transmission rates change at pre-specified dates in EpiBeds. Some of these correspond to large policy changes, so can be added as soon as we expect them to become visible in the data (7 days after policy change). Others correspond to noticeable changes in the gradient of the admissions data on a log scale. These are only added once there is sufficient data to know that there has been a genuine change in the trend, which generally occurs 15 days after the change point. Finally, when fitting EpiBeds, a final change point is added three weeks prior to the last data point, provided there is no other change point in this window. Therefore, when generating the scenarios presented in Section 2.4, the changes points are specified as indicated in Table M.3.

**Table M.3:** *Dates of transmission rate changes for the 19 scenarios considered in Section 2.4. For the 15/04/20 forecast, a change point is not added three weeks before the last data point since this would be one day after lockdown, and therefore risks overfitting the data.*

| Forecast start date | 1 | 2 | 3 | 4 | 5 | 6 | 7 | 8 | 9 |
|---|---|---|---|---|---|---|---|---|---|
| **01/04/20** | 13/03/20 | 24/03/20 | - | - | - | - | - | - | - |
| **15/04/20** | 13/03/20 | 24/03/20 | - | - | - | - | - | - | - |
| **01/05/20** | 13/03/20 | 24/03/20 | 09/04/20 | - | - | - | - | - | - |
| **15/05/20** | 13/03/20 | 24/03/20 | 11/04/20 | 24/04/20 | - | - | - | - | - |
| **01/06/20** | 13/03/20 | 24/03/20 | 11/04/20 | 10/05/20 | - | - | - | - | - |
| **15/06/20** | 13/03/20 | 24/03/20 | 11/04/20 | 25/05/20 | - | - | - | - | - |
| **01/07/20** | 13/03/20 | 24/03/20 | 11/04/20 | 09/06/20 | - | - | - | - | - |
| **15/07/20** | 13/03/20 | 24/03/20 | 11/04/20 | 24/06/20 | - | - | - | - | - |
| **01/08/20** | 13/03/20 | 24/03/20 | 11/04/20 | 10/07/20 | - | - | - | - | - |
| **15/08/20** | 13/03/20 | 24/03/20 | 11/04/20 | 25/07/20 | - | - | - | - | - |
| **01/09/20** | 13/03/20 | 24/03/20 | 11/04/20 | 10/08/20 | - | - | - | - | - |
| **15/09/20** | 13/03/20 | 24/03/20 | 11/04/20 | 15/08/20 | 25/08/20 | - | - | - | - |
| **01/10/20** | 13/03/20 | 24/03/20 | 11/04/20 | 15/08/20 | 09/09/20 | - | - | - | - |
| **15/10/20** | 13/03/20 | 24/03/20 | 11/04/20 | 15/08/20 | 06/09/20 | 24/09/20 | - | - | - |
| **01/11/20** | 13/03/20 | 24/03/20 | 11/04/20 | 15/08/20 | 06/09/20 | 14/10/20 | - | - | - |
| **15/11/20** | 13/03/20 | 24/03/20 | 11/04/20 | 15/08/20 | 06/09/20 | 14/10/20 | 05/11/20 | - | - |
| **01/12/20** | 13/03/20 | 24/03/20 | 11/04/20 | 15/08/20 | 06/09/20 | 14/10/20 | 05/11/20 | 09/11/20 | - |
| **15/12/20** | 13/03/20 | 24/03/20 | 11/04/20 | 15/08/20 | 06/09/20 | 14/10/20 | 05/11/20 | 18/11/20 | 24/11/20 |
| **01/01/21** | 13/03/20 | 24/03/20 | 11/04/20 | 15/08/20 | 06/09/20 | 14/10/20 | 05/11/20 | 18/11/20 | 09/12/20 |

## Data and Materials

Code for simulating the EpiBeds model is available at: https://github.com/OvertonC/EpiBeds. The data used were provided through a data sharing agreement, and unfortunately cannot be provided. Data similar to the SITREP and CPNS sources are available at, though with slightly different data definitions, are available at: https://coronavirus.data.gov.uk/details/download.

## Author contributions

| |
|---|
| **Conceptualization:** CO, LP, KL |
| **Literature review:** HS, AN, KL |
| **Data Curation:** HS, AN, CO, LP, KL, FS |
| **Formal Analysis:** CO, HS, LP, KL, JB, FP, FS |
| **Methodology:** CJ, CO, KL, LP, FS, JB, FP, TH, IH |
| **Software:** CO, LP, JB, FP, FS |
| **Visualization:** CO, KL, LP, HS |
| **Writing – Original Draft Preparation**: CO, HS, AN, FS, KL, LP, CF |
| **Writing – Review & Editing:** CO, HS, AN, FS, KL, LP |

## Funding

L.P., H.B.S. and C.E.O. are funded by the Wellcome Trust and the Royal Society (grant no. 202562/Z/16/Z). F.P. was funded through a PhD scholarship at the University of Manchester. J.B. was supported by a Wellcome Trust Four-Year PhD Studentship in Basic Science (219992/Z/19/Z). C.J. is funded by the MRC (MR/V038613/1), EPSRC (EP/W011840/1, EP/R018561/1), and Wellcome (UNS73114). T.A.H. is supported by the Royal Society (grant no. INF\R2\180067). I.H. is supported by the National Institute for Health Research Health Protection Research Unit (NIHR HPRU) in Emergency Preparedness and Response and the National Institute for Health Research Policy Research Programme in Operational Research (OPERA). C.E.O., L.P., I.H., T.A.H. and F.S. are supported by the UKRI through the JUNIPER modelling consortium [grant number MR/V038613/1]. K.A.L. is funded by the Wellcome Trust and The Royal Society (107652/Z/15/Z).

## Acknowledgements

The authors would like to thank colleagues in SPI-M-O and JUNIPER consortium for various discussions around hospital modelling and forecasting.

## Conflict of Interest

The authors declare that they have no conflict of interest.

# Appendices

## Appendix A Input parameters for short-term forecasting evaluation

When generating the short-term forecasts in Section 2.4, we required point estimates for the length of stay and priors for the probability of death on ICU, using data available at the time of each forecast. We generated estimates using data available at the end of each month, from the end of March 2020 onwards, which were used for any forecasts generated in the subsequent month. The resultant estimates are given in Tables A.1 and A.2.

In Table A.1, we present the monthly length of stay estimates, which represent the best estimates available up to the end of the given month. Looking at the estimates over time, we observe that although the method (Section M.2) attempts to compensate for the truncated tail observations, it can take a while for it to receive enough information to provide reliable estimates: moving from March to May, the mean lengths of stay gradually increase, as more information arrives with which to parameterise the tail of the distribution. This is to be expected, since in March and April, there was a maximum of 31 and 61 days, respectively, for the observed lengths of stay, so any tail observations longer than this would not be identified, and therefore EpiBeds does not have sufficient information to correctly adjust for the truncated tail. From May to August, the estimates are mostly stable. A similar pattern occurs as we move from September to December, whereby it takes a couple of months for the estimates to stabilise.

In Table A.2, we present the monthly prior estimates for $p_D$. Initial $p_D$ estimates were very high, likely driven by patients that recover being more affected by right-censoring in their outcomes, since the length of stay to discharge is longer than the length of stay to death. In the second wave, initial $p_D$ estimates were low. At this point, the length of stay to death was longer than to discharge, so here the deaths are more affected by the right truncation. Additionally, during the second wave the age distribution of admissions changed as the epidemic progressed, resulting in older patients being admitted, who had a higher rate of mortality in ICU.

**Table A.1:** *Point estimates for the mean length of stay using data until the end of each month. From March to August, this uses all data from 01/03/2020. From September to December, this uses all data from 01/08/2020.*

| Length of stay | March | April | May | June | July | August | September | October | November | December |
| --- | --- | --- | --- | --- | --- | --- | --- | --- | --- | --- |
| Hosp to ICU | 1.46 | 2.1 | 2.49 | 2.65 | 2.86 | 2.79 | 1.82 | 2.62 | 2.53 | 2.70 |
| ICU to death | 7.61 | 9.96 | 11.10 | 11.4 | 11.55 | 11.84 | 9.17* | 14.71 | 14.88 | 15.33 |
| ICU to monitoring | 5.58 | 11.25 | 16.5 | 16.63 | 16.10 | 15.93 | 6.51 | 6.86 | 7.46 | 8.57 |
| Monitoring to recovery | 6.91 | 6.72 | 9.29 | 10.76 | 11.31 | 11.85 | 4.32 | 6.59 | 6.27 | 6.45 |

| | | | | | | | | | |
|---|---|---|---|---|---|---|---|---|---|
| **Hosp to recovery (no ICU)** | 5.13 | 8.19 | 8.91 | 9.20 | 9.24 | 9.37 | 6.89 | 10.04 | 10.07 | 10.02 |
| **Hosp to death (no ICU** | 8.09 | 7.79 | 8.45 | 8.68 | 8.86 | 8.93 | 5.20 | 10.46* | 13.37 | 12.16 |

**Table A2:** *Prior estimates for the probability of dying on ICU, using data until the end of each month. From March to August, all data from 01/03/2020 is used. From September to December, all data from 01/08/2020 is used.*

| $p_D$ | March | April | May | June | July | August | September | October | November | December |
|---|---|---|---|---|---|---|---|---|---|---|
| mean | 0.46 | 0.42 | 0.372 | 0.355 | 0.358 | 0.357 | 0.159 | 0.252 | 0.293 | 0.305 |
| standard deviation | 0.0335 | 0.018 | 0.0215 | 0.018 | 0.018 | 0.017 | 0.033 | 0.026 | 0.0195 | 0.0165 |